RESEARCH ARTICLE

JASIST WILEY

# FAIR: Fairness-aware information retrieval evaluation

Ruoyuan Gao[1] | Yingqiang Ge[1] | Chirag Shah[2]

[1]Department of Computer Science, Rutgers University, Piscataway, New Jersey, USA

[2]Information School, University of Washington, Seattle, Washington, USA

**Correspondence**
Ruoyuan Gao, Department of Computer Science, Rutgers University, Piscataway, NJ, USA.
Email: gaoruoyuan@gmail.com

**Abstract**

With the emerging needs of creating fairness-aware solutions for search and recommendation systems, a daunting challenge exists of evaluating such solutions. While many of the traditional information retrieval (IR) metrics can capture the relevance, diversity, and novelty for the utility with respect to users, they are not suitable for inferring whether the presented results are fair from the perspective of responsible information exposure. On the other hand, existing fairness metrics do not account for user utility or do not measure it adequately. To address this problem, we propose a new metric called FAIR. By unifying standard IR metrics and fairness measures into an integrated metric, this metric offers a new perspective for evaluating fairness-aware ranking results. Based on this metric, we developed an effective ranking algorithm that jointly optimized user utility and fairness. The experimental results showed that our FAIR metric could highlight results with good user utility and fair information exposure. We showed how FAIR related to a set of existing utility and fairness metrics and demonstrated the effectiveness of our FAIR-based algorithm. We believe our work opens up a new direction of pursuing a metric for evaluating and implementing the FAIR systems.

## 1 | INTRODUCTION

When a user initiates a search request, the information retrieval (IR) system must efficiently explore the search space and return a set of relevant results. Since relevance is often considered the most important metric of the system, the result set may not guarantee enough diversity coverage. Items in the largest or the most prominent subtopic/aspect group may take up most of the search results, thus leaving little space for the minority aspects to surface. Consequently, it may take a user much more effort to reach an item of a minority aspect of a query topic, leading him/her to a partial—and often very skewed—discovery of information. Furthermore, because users frequently only click on the top results, a ranking or recommendation algorithm that takes user feedback into account, will continue putting those same items at the top, creating a positive feedback loop and a vicious cycle of unfairness (Baeza-Yates, 2018; Joachims et al., 2017). In this article, we study fairness in open-domain web search with respect to subtopics and minority aspects of a topic. Similar to (Gao & Shah, 2020), we refer to each subtopic or aspect as a group and define fairness as a subjective moderation of the ratio between different groups. This ratio is determined by a policy maker of the IR service provider, such that whether the results are fair with respect to subtopics is purely defined from the service provider's point of view and not depend on users. For example, benefits and risks can be considered as two subtopics of the topic "coffee health," or discussions focusing on different genders can be considered different aspect groups. IR service providers may determine that the query "who commits crimes" returning equal number of results on people of different genders is fair, or they may determine that the results reflecting the gender proportion in criminal records is fair. The lack of fairness here stems from data as well as algorithms, and is further reinforced by users' indifference or ignorance with regards to how a system ranks and presents such information.

A diversification framework aims to maximize the coverage of topical aspects. This helps increase diversity and in turn, fairness, assuming that such fairness requires certain exposure of minority subtopics (E. Celis, Keswani, et al., 2018; L. E. Celis et al., 2019; Graells-Garrido & Lalmas, 2014). However, the goal of diversity and fairness is fundamentally different (Gao & Shah, 2020). On the one hand, system utility such as diversity serves a user's interest directly, so that users with different information needs can find relevant information. Fairness, on the other hand, comes from a social perspective with the purpose of social responsibility. It moderates the exposure of information so that different information and resources get fair chances to receive users' attention. The optimization goal of diversity is to cover as many subtopics as possible for users. Fairness, in contrast, constraints how much exposure each subtopic should receive on the IR system side. Because of the difference in purpose and target, fairness and diversity in IR can sometimes become competing factors (Diaz et al., 2020; Gao & Shah, 2020). For example, a system can provide a poor exposure of a minority subtopic but still be fair (Demartini & Siersdorfer, 2010). Subsequently, existing diversification metrics are not sufficient to serve as the fairness metric we address in this article.

While existing evaluation metrics are able to examine individual factors of fairness (Geyik et al., 2019; Kuhlman et al., 2019), relevance (Moffat & Zobel, 2008), diversity, and novelty (Clarke et al., 2008), they are limited in capturing the overall performance combining all factors. First off, the evaluation should capture the algorithm's capability of eliminating bias and achieving fairness. Second, it is not enough to simply evaluate an algorithm's effectiveness on achieving fairness. As the fundamental functionality of an IR system, utility must also be evaluated. Therefore, the evaluation must capture the connections between fairness and utility. There is no standard metric that is able to capture such trade-offs, which poses a challenge when comparing different algorithms. As a result, an evaluation metric that accounts for both utility and fairness will be better suited for fairness ranking.

The lack of a unified metric is not only inconvenient from the perspective of evaluation, but also leads to difficulties in fairness optimization. Empirically, it has been shown that simply optimizing for system utilities lead to biased results (Diaz et al., 2020; Singh & Joachims, 2018), which is exactly the algorithmic bias problem with existing systems that aim for optimal utility. Meanwhile, optimizing for fairness often leads to a decrease in system utility (Gao & Shah, 2020; Mehrotra et al., 2018; Zehlike et al., 2017). Therefore, in the case that both utility and fairness are important, the algorithm that jointly optimizes both factors may achieve better results. Based on our proposed metric, we are able to construct a randomized fairness ranking algorithm that optimizes for fairness as well as utility while capturing the balance between these two factors. To sum up, our work makes the following contributions:

- We propose a novel fairness-aware evaluation metric for the ranking results that encodes fairness and user utility. This metric offers a new perspective and foundation of evaluating fairness-aware ranking results and can be generalized to many IR applications. Our experiments show that this metric is able to highlight results and algorithms that are optimal for both fairness and utility dimensions, while penalizing algorithms that are skewed toward one dimension.
- We demonstrate the usefulness of our new metric through experiments on various datasets. The experimental results indicate the robustness of our metric and its strong capability of measuring both utility and fairness.
- We develop a ranking algorithm that optimizes for the proposed fairness-aware metric. Our proposed algorithm does not require the gold standard relevance judgment, which is often unavailable in practice. We demonstrate the effectiveness of the proposed algorithm through a detailed case study with respect to improving fairness while closely approximating the ideal utility.

The rest of the article is organized as follows. We provide an overview of related work. Next, we describe details of our proposed metric and optimization algorithm. We then present our experiments along with detailed analyses. Finally, in the last two sections, we discuss the limitations and implications and conclude our work.

## 2 | RELATED WORK

Our work is closely related to the field of standard evaluation metrics in IR, bias and fairness measurements situated in the IR context, and fairness ranking algorithms that address the trade-offs between fairness and retrieval utility via post-processing and randomization.

### 2.1 | Evaluation metrics in IR

Traditional evaluation measures in IR focus on the quality and utility of the search and recommendation results from the perspective of the user. Metrics that are based on precision and recall measure how well the whole results are relevant to the user's information need. Examples of such metrics include *Mean Average Precision* (MAP), *Discounted Cumulative Gain* (DCG), and *Rank-Biased Precision* (RBP; Moffat & Zobel, 2008) which also considered the position of the relevant result. Diversity and novelty metrics aim to satisfy different user needs that arise from the ambiguity or multiple aspects of the query. For instance, $\alpha$-nDCG (Clarke et al., 2008) rewards a candidate document for being relevant to a large number of yet novel aspects based on previously selected documents. *Intent-Aware* (IA) metrics (Agrawal et al., 2009; e.g., *Intent-Aware Expected Reciprocal Rank*, ERR-IA; Chapelle et al., 2009) assume a probability distribution over user intents for a query. Intent-aware metrics aim to satisfy the "average user" more than the long tail users and also captures diversity and novelty. The classic IR metrics measure the utility of the system at serving the user's information need. They cannot measure fair information exposure.

## 2.2 | Fairness notions and measures

Fairness in IR have been often studied from the perspective of information exposure regarding sensitive attributes such as gender and race (Singh & Joachims, 2018). Different items should receive equal exposure, or exposure proportional to their utilities or impacts, depending on which exposure distribution is considered fair by the system designer. Our work draws from previous works (L. E. Celis et al., 2019; Diaz et al., 2020; Gao & Shah, 2020) to view topical aspects as sensitive attributes and focuses on topical group fairness. This enables us to situate fairness in the diversity context and connect it with standard IR metrics.

More and more works have attempted to propose fairness metrics, according to various constraints such as distance and ratio between the proportion of a protected attribute and the overall attribute proportion (K. Yang & Stoyanovich, 2017), pairwise comparisons regarding utility and prediction errors (Beutel et al., 2019; Kuhlman et al., 2019; Yao & Huang, 2017b), exposure distributions against the desired distribution (Geyik et al., 2019; K. Yang & Stoyanovich, 2017). Several studies have compared existing fairness metrics (Chouldechova, 2017; Garg et al., 2020; Hinnefeld et al., 2018; Raj et al., 2020; Sapiezynski et al., 2019). Unlike metrics that only capture fairness, our work focuses on incorporating fairness into a unified metric that also accounts for standard IR metrics. As a design of a new evaluation metric, our fairness portion is not limited to a single fairness definition, but can adopt distribution-based, ratio-based, or error-based fairness metrics. We show that in IR, an integrated metric is able to offer a better balance between utility and fairness.

The closest work to us is Liu et al. (2020), which proposed to combine accuracy and fairness in the reward function for reinforcement learning, aiming to maintain the accuracy-fairness trade-off in interactive recommendation. Yet the reward function is more focused on optimization approach rather than an evaluation metric that can be directly applied to measure a given rank list. We make the explicit contribution to a metric for evaluating fairness-aware ranking. Another closely related work is Diaz et al. (2020) which proposed the *expected exposure* to capture fairness as well as the user models from ERR and RBP. Rather than evaluate stochastic rankings, our work is situated in static ranking evaluation and accommodates various use cases that do not assume stochastic ranking models.

## 2.3 | Fairness ranking algorithms

Fairness ranking has been investigated in many works as an optimization problem. A majority of the fairness ranking algorithms aim to optimize the system utility while satisfying a set of fairness constraints (L. E. Celis, Straszak, & Vishnoi, 2018; Singh & Joachims, 2018; Zehlike et al., 2017). Based on an existing ranking or estimation of item utility, fairness can be improved by re-ranking the items based on the fairness constraints (Gao & Shah, 2020), or by applying fairness as a regularizer in the objective function (Asudeh et al., 2019; Beutel et al., 2019; L. E. Celis et al., 2019; Mehrotra et al., 2018; Wan et al., 2020; Yao & Huang, 2017a). In practice, optimizing for one metric often leads to the decrease in another (Diaz et al., 2020; Mehrotra et al., 2018; Singh & Joachims, 2018). In theory, the correlation between fairness and system utility depends on the characteristics of the data (Gao & Shah, 2019). Therefore, algorithms for constrained maximization may not lead to the optimal fairness and utility. Similar to our work, Mehrotra et al. (2018) and L. E. Celis et al. (2019) proposed randomized algorithms for jointly optimizing fairness and relevance. Our algorithm directly optimizes the proposed integrated metric. It benefits from being able to capture the balance between utility and fairness inferred from the evaluation metric. It is designed for fairness ranking and thus can be generalized to the recommendation setting.

## 3 | FAIRNESS-AWARE IR (FAIR) METRIC

We propose the *F*airness-*A*ware *IR* (FAIR) metric that unifies fairness with standard IR utility metrics. Previous works on fairness metrics do not account for the precision, diversity, novelty, or search intent. As a result, such metrics cannot capture the connections between system utility and fairness, and cannot serve as a unified metric that evaluates the system's performance. In addition, algorithms that are designed to directly optimize such fairness metrics cannot guarantee, and often decrease the optimal utility (Diaz et al., 2020; Gao & Shah, 2019; Mehrotra et al., 2018; Zehlike et al., 2017). While algorithms that jointly optimize for both fairness and system utility attempt to balance between the two, they are still problematic due to the fact that they treat fairness and utility separately and as competing factors. As discussed earlier, fairness and utility are not necessarily orthogonal. In fact, they can benefit each other in many cases. FAIR treats utility and fairness as factors of the unified metric. We carefully design FAIR this way to avoid making assumptions of the relationship between utility and fairness. If the relationship is orthogonal, then optimizing FAIR means balancing the trade-offs between them; if not, then optimizing FAIR means optimizing both of them.

## 3.1 | Overall design

We use the position discounted KL-divergence to demonstrate the fairness portion. In practice, this portion can be replaced with other fairness metrics such as ratio-based (e.g., minSkew; Geyik et al., 2019) and error-based metrics (e.g., Rank Parity; Kuhlman et al., 2019). Formally, for a rank of $k$ documents,

$$\text{FAIR} = \frac{1}{M} \sum_{i=1}^{k} \frac{\text{IRM}_i}{d_{KL}(D_{r^i} \| D^*) + 1}, \quad (1)$$

where $d_{KL}(D_{r^i} \| D^*)$[1] is the KL-divergence between the subtopic distribution in the top-$i$ ranking $D_{r^i}$ and the desired subtopic distribution $D^*$. $\text{IRM}_i$ is a standard IR metric measured at rank $i$. To capture the precision, IRM can be replaced with metrics like MAP and RBP. To compute the intent-aware FAIR, we can replace IRM with a metric like the ERR-IA. To emphasize diversity and novelty, we can replace IRM with a metric like *Novelty- and Rank-Biased Precision* (NRBP; Clarke et al., 2009) and $\alpha$-nDCG. In other words, FAIR aims to serve as a general fairness-aware metric, where the specific focus on intent or diversity is captured by the IRM term. $M$ is a normalization term that aims to bring the entire metric into the range [0,1]. Therefore, $M$ should be an ideal score that the metric can achieve, and we divide the actual score by the ideal score to get a normalized score. As we will describe later, an example of $M$ can be the ideal discounted cumulative gain (*IDCG*) to normalize *DCG*. Note that because $d_{KL} \geq 0$, we have $d_{KL} + 1 \geq 1$. When there is no bias as quantified by the KL-divergence, or when bias is not factored into the evaluation metric, FAIR reduces to the standard normalized IR metric.

## 3.2 | Encoding utility

To compute FAIR, the first step is to compute the IRM. Here we demonstrate how to design a diversity- and novelty-based fairness-aware metric by adopting the model from $\alpha$-nDCG. In this model, a query is considered to be ambiguous or containing multiple aspects. Thus, if we view each of the aspects as an information need, we can quantify the diversity as the number of information needs covered, and novelty as the amount of new information. Let $D$ be the collection of documents for a query. $J(d,a)$ denotes the relevance judgment that the document $d \in D$ contains the diversity aspect $a$ which is a information need among the set of aspects $A$. $d_i$ denotes the document at rank $i$. Due to the different user needs, Clarke et al. (2008) proposed to compute the probability that $d_i$ is relevant given the user need $a \in A$ as

$$P(R_i = 1 | a, d_1, d_i, ..., d_i) \approx \prod_{a \in A} J(d_i, a)(1-\alpha)^{r_{a,i-1}}, \quad (2)$$

where

$$r_{a,i-1} = \sum_{j=1}^{i-1} J(d_j, a) \quad (3)$$

is the sum of relevance judgment scores up to rank $i - 1$. The information gain of $d_i$ is then defined as

$$G[i] = \sum_{a \in A} J(d_i, a)(1-\alpha)^{r_{a,i-1}}. \quad (4)$$

The cumulative gain up to rank $k$, denoted by $CG[k]$, is the sum of $G[i]$, $i = 1, 2, ..., k$. To account for the position bias, we discount $CG[i]$ by a factor of $\log_2(i+1)$ to get the discounted cumulative gain $DCG[i]$. Therefore, we have

$$DCG[k] = \sum_{i=1}^{k} \frac{G[i]}{\log_2(i+1)}. \quad (5)$$

In this model, the relevance and information gain is more important at higher ranked positions. *IDCG* can be approximated using a greedy algorithm as discussed in Clarke et al. (2008). We divide *DCG* by *IDCG* to get the score of α-nDCG.

## 3.3 | Encoding fairness

Previous KL-divergence based fairness metrics (e.g., Geyik et al., 2019; K. Yang & Stoyanovich, 2017) discount the KL-divergence distance with position bias using $\frac{d_{KL}(D_{r^i}\|D^*)}{\log_2(i+1)}$. Due to the fact that the higher the $d_{KL}$, the lower the fairness, such metrics encourage fairness at lower ranks. For example, given the same KL-divergence $d_{KL}(D_{r^i}\|D^*) = d_{KL}(D_{r^j}\|D^*)$ at a higher rank position $i$ and a lower rank $j$, $i<j$, $\frac{d_{KL}(D_{r^j}\|D^*)}{\log_2(j+1)}$ is apparently better than $\frac{d_{KL}(D_{r^i}\|D^*)}{\log_2(i+1)}$ as we wish to minimize the distribution difference. In practice, the higher ranked documents tend to receive more attention and thus have more impact on users. Therefore, the position discount factor $\log_2(i+1)$ is designed to diminish the utility at the lower rank position, so that it encourages highly useful documents to be placed at the higher rank. In the case of fairness, it is also desirable that the system presents the resources more fairly at the top ranked positions because of the impact of position bias. In our proposed metric FAIR, we address this problem by taking the reciprocal of $d_{KL}$ and applying the position discount on the reciprocal of $d_{KL}$. Note that if the position discount factor is already encoded in the IRM, there is no need to repeat this discount for fairness. With the α-nDCG example, we can replace the $IRM_i$ and $M$ accordingly and rewrite Equation (1) as

$$\text{FAIR} = \frac{1}{IDCG} \sum_{i=1}^{k} \frac{1}{\log_2(i+1)} \cdot \frac{G[i]}{d_{KL}(D_{r^i}\|D^*) + 1}. \quad (6)$$

This metric encodes the diversity and novelty through $G[i]$ as defined in Equation (4), accounts for fairness through the reciprocal of KL-divergence $\frac{1}{d_{KL}}$, and addresses the position bias by the discounting factor $\log_2(i+1)$ with respect to rank position $i$. In other words, the diversity- and novelty-based relevance get discounted by the bias at each rank position. Consequently, the fairer the ranking, the better FAIR approximates the optimal α-nDCG. This gives us a relevance metric that not only accounts for diversity and novelty, but also is fairness-aware.

FAIR is designed to be a general evaluation metric for fairness-aware IR systems. This means it is flexible to account for various standard IR metrics. That being said, we must be careful when integrating any IR metrics into FAIR. (1) When using the summation form defined in Equation (1), $IRM_i$ must be a utility score for rank $i$ only. It cannot be an integrated score accumulated from previous ranks. For example, with α-nDCG, we used the information gain of a document at the $i$-th rank $G[i]$, instead of the cumulative gain $CG[i]$ which was a sum of gains from rank 1 to rank $i$. If we adopt the model based on the RBP metric, which assumes that the user will look at the $i$-th ranked document $d_i$ with probability $p$,

$$\text{RBP} = (1-p)\sum_i J(d_i)p^{i-1}, \quad (7)$$

then $IRM_i$ should be replaced by $J(d_i)p^{i-1}$, where $J(d_i)$ is the relevance judgment for $d_i$. However, if the IR metric is not an accumulation of scores at each rank, for example, precision, we may want to rewrite FAIR as $\frac{IRM}{d_{KL}(D_{r^i}\|D^*)+1}$. (2) The position discount can be different and may or may not be necessary depending on the user model captured by the IR metric. When the IR metric includes position discount, we should avoid additional position discounts associated with fairness. For example, with RBP, the discount factor $\log_2(i+1)$ in Equation (6) should be replaced by $p^{i-1}$ while $G[i]$ is replaced by $J(d_i)$. (3) While KL-divergence has been used as a measure for bias in many previous works, the $d_{KL} + 1$ may be replaced by other metrics such as the distance between two fairness representation vectors (Chen et al., 2018; Equation 2), the squared error between the expected exposure and the desired exposure (Diaz et al., 2020; Equation 1). Note that for (1), the intuition of using $IRM_i$ when possible is that, to best capture both utility and fairness across the entire ranking list, we would like to be precise about the combined score at each rank position. It may seem that, at each rank position $i$, $d_{KL}$ computes the KL-divergence for the top-$i$ items, thus one may be tempted to use the cumulative utility IRM up to rank $i$ instead of $IRM_i$ as well. However, $d_{KL}$ at $i$ is not an accumulation of KL-divergence scores. Instead, it is a score directly computed based on the group distributions in top-$i$ when the $i$-th item is added. It reflects the fairness status *at* rank $i$. This is consistent with Equation (4), where $IRM_i = G[i]$ is directly computed on the information gain with respect to top-$(i-1)$ when the $i$-th item is added. $IRM_i$ reflects the information gain *at* rank $i$.

## 3.4 | Capturing different fairness notions

Different fairness notions can be achieved by varying the group distribution. Geyik et al. (2019) explained how to design the desired distribution such that the desired distribution represented the *equal opportunity* (Hardt et al., 2016) and *demographic parity* (Singh & Joachims, 2018) criteria. We refer readers to Geyik et al. (2019) for details on how to design the desired distribution to capture different fairness notions. Basically, parity-based group fairness requires that every subtopic in the ranking receives the equal opportunity to be exposed. In static ranking, this means equal number of subtopics should be presented in the ranking (e.g., equal number of results on benefits and risks of coffee). When utility matters, some subtopics may contain more relevant items. In this case, we may desire the *disparate treatment* fairness (Singh & Joachims, 2018), that is, the proportion of items from each subtopic should be proportional to the subtopic's average relevance. To reflect the distribution of groups in the data, we can implement *demographic parity* fairness (Singh & Joachims, 2018) by setting the desired distribution to be the equal to the distribution of subtopics among all candidate documents (e.g., 70% benefits and 30% risks of coffee).

## 4 | FAIRNESS-AWARE RE-RANKING ALGORITHM

In this section, we propose a fairness-aware re-ranking algorithm *FAIR $\varepsilon$-greedy* (Algorithm 1) to optimize for FAIR metric in the top-$k$ ranking. This algorithm is based on the idea of $\varepsilon$-greedy algorithm that aims to balance between fairness and utility (L. E. Celis et al., 2019; Gao & Shah, 2020; Mehrotra et al., 2018) in search and recommendations. In this algorithm, we utilize our proposed FAIR metric to customize the balance between fairness and the fairness-aware system utility. Again, we use the $\alpha$-nDCG-based FAIR (Equation 6) for subtopical group fairness as an example. We can also use the RBP-based FAIR to design the algorithm. The only change that needs to be made is to replace the $\alpha$-nDCG formula with RBP formula (Equation 7). At each rank $i$, with probability $1 - \varepsilon$, we optimize for $\frac{G[i]}{d_{KL}}$, the diversity and novelty gain discounted by bias and rank position. With probability $\varepsilon$ we explore documents that minimize the KL-divergence $d_{KL}$.

We note two special cases with this algorithm. The first one is when $\varepsilon = 0$, this algorithm becomes a simple greedy algorithm that always optimizes for the FAIR metric. Given the trade-off between $\alpha$-nDCG and $d_{KL}$ defined in Equation (6), 0-greedy optimizes both metrics while maintaining this trade-off. The second case is when $\varepsilon = 1$, this algorithm becomes a simple greedy algorithm that minimizes the KL-divergence at each rank. In this case, $\alpha$-nDCG is not considered although increasing fairness may benefit diversity for some datasets (Gao & Shah, 2020).

The FAIR $\varepsilon$-greedy algorithm is a ranking algorithm over the retrieved items or the collection of candidate items for recommendation. Hence it does not require the gold standard relevance judgment labels. When the relevance judgment labels are unavailable, we can reference the default ranking provided by a retrieval system. We assume that the top ranked items in the default ranking are all relevant and the higher ranked items are more relevant than the lower ranked ones. In the case of recommender systems, we assume that the top recommended items in the default recommendation set are higher in utility (e.g., click-through-rate, probability of being purchased).

## 5 | EXPERIMENTS

In this section, we describe our experimental setups and results. We first examine the relationship between our FAIR

---

**ALGORITHM 1    FAIR $\varepsilon$-greedy**

**Input**: $k \geq 1$, $\varepsilon \in [0,1]$, a desired subtopic distribution $D^*$, an initial ranking list with the corresponding utility value $R_0$

**Output**: $R_k$

   initialize $R_k = [\ ]$
   **for** $i \in [1:k]$ **do**
      **with probability** $1 - \varepsilon$:
         candidates $= \{d : \mathrm{argmax}_d \frac{G[i]}{d_{KL}(D_{r^i} \| D^*)},\ d \in R_0\}$
         nextDoc $= \mathrm{argmin}_d d_{KL}(D_{r^i} \| D^*), d \in$ candidates
      **with probability** $\varepsilon$:
         candidates $= \{d : \mathrm{argmin}_d d_{KL}(D_{r^i} \| D^*),\ d \in R_0\}$
         nextDoc $= \mathrm{argmax}_d G[i], d \in$ candidates
      $R_k[i] = $ nextDoc
   **end for**

metric and commonly used utility and fairness metrics. Then, we present a case study to further understand how FAIR changes when different metrics are optimized. Lastly, we evaluate the performance of our FAIR $\varepsilon$-greedy algorithm.

## 5.1 | Comparison between FAIR and other metrics

Here we investigate the FAIR metric on different datasets and ranking algorithms. Given a dataset and a ranking algorithm, we compute the utility scores, fairness scores, as well as the FAIR scores. Through comparisons within a dataset and between various datasets, we show how FAIR captures the utility and fairness dimensions at the same time. We also include the correlation analysis to understand how FAIR correlates with other metrics.

### 5.1.1 | Datasets

Three datasets were used in our experiments. Details on how subtopics are defined or measured are described in the references for each dataset.

- Google Search: This is the full dataset used by Gao and Shah (2020). These data contain 100 queries and 7,410 webpages crawled from a snapshot of Google Web search. For each query, the first 100 search results in their original ranking order returned by Google are provided, along with the binary relevance judgment with respect to the query. In this dataset, there are two suptopical groups per query. Each document belongs to exactly one subtopical group.
- ClueWeb09: ClueWeb09 is the dataset used in the TREC 2009 Web Track,[2] commonly used in the diversity-related retrieval tasks. These data contain roughly 1 billion web pages comprising approximately 25TB of uncompressed data in multiple languages, crawled from the Web during January and February 2009. Fifty topics with 3–8 subtopics per topic, along with the binary relevance judgment with respect to the subtopics, are provided. In this dataset, one document can be relevant to none or one or multiple subtopics. Same as the TREC 2009 Web Track diversity task, a document is considered relevant to a topic if it was relevant to any of the subtopics.
- MovieLens: MovieLens (Harper & Konstan, 2015) is a public benchmark dataset commonly used in the recommendation literature, especially in recent recommendation works concerning fairness (Ge et al., 2021; T. Yang & Ai, 2021). We choose MovieLens-20 M which includes over 20 million user transactions (user id, item id, rating, timestamp, etc.) between January 9, 1995 and March 31, 2015. The same as Morik et al. (2020), fairness is considered with respect to production companies. We selected the two production companies with the most movies in the dataset, MGM and Warner Bros. After removing movies with very few ratings, we obtained a partially filled ratings matrix with 10,000 users, 100 movies, and 46,515 user-item interactions, with sparsity around 95.35%.

### 5.1.2 | Fairness constraint and algorithms

For fairness constraint, we examined the *demographic parity* fairness by setting the desired distribution $D^*$ to be the group distribution in the entire document/item collection for a query/user. Note that our metric can incorporate different fairness notions by setting $D^*$ to be of different distributions. To investigate the FAIR scores in different application scenarios, we adopted the following algorithms in our experiments: Google's original ranking (Google) for Google Search data, DetCons for Google Search and ClueWeb09, and FOE for MovieLens. Maximizing Greedy Conservative Mitigation Algorithm (DetCons; Geyik et al., 2019) is one of the state-of-the-art fairness ranking algorithm. DetCons assumes the items can be ordered in the descending order of relevance scores. On Google Search data, we used the original Google ranking as the ordering of documents. On the ClueWeb09 data, we used the gold standard relevance judgment for this ordering. Fairness of exposure in ranking (FOE; Singh & Joachims, 2018) is a general framework that employs probabilistic rankings and linear programming to compute the utility-maximizing ranking under a whole class of fairness constraints. It is one of the state-of-the-art reranking framework based on group fairness constraints. In our experiments, FOE was used to maximize the user-item rating scores given the demographic parity fairness constraint. We modified FOE to accommodate the recommendation task through replacing the query-document relevance scores with user-item rating scores. We did not compare the algorithms that jointly optimize for fairness and relevance in the recommendation setting (L. E. Celis et al., 2019; Mehrotra et al., 2018) as these two algorithms are not directly comparable to our ranking algorithm. Instead, they focus on improving the fairness and relevance on a set of items that do not care about the ranking between items in the same set. We did not include BM25 as a baseline because BM25 has shown to be weaker than the original Google baseline (Gao & Shah, 2020).

## 5.1.3 | Evaluation metrics

We employed the following metrics: *FAIR, nDCG, RBP, KL-divergence (KL), nDRKL*. We report the top-$k$ scores for each metric with $k = 10, 20, 50$ to account for various rank positions. nDCG and RBP represent the utility functions that measure the system's capability of retrieving relevant results for the user, taking into account of the impact of rank position. For fairness evaluation, we employed the KL as a direct indicator of the how different the group distributions in the ranking were from the desired distributions. We also employed a modified nDKL (Geyik et al., 2019; K. Yang & Stoyanovich, 2017), termed nDRKL to measure the rank-biased unfairness. The Normalized Discounted KL-divergence (nDKL) discounts the bias at each rank position, so that the higher the rank, the lower the discount on the bias. This metric highlights the highly biased results at earlier rank. Yet for the purpose of achieving fairness, we wish to highlight the highly fair results at earlier rank. Therefore, we propose the *Normalized Discounted Reciprocal of KL-divergence* (nDRKL) metric. nDRKL is modified based on nDKL by replacing the KL-divergence with its reciprocal at each rank position. Specifically, for the top-$k$ ranking,

$$\text{nDRKL} = \frac{1}{Z}\sum_{i}^{k} \frac{1}{\log_2(i+1)} \cdot \frac{1}{d_{KL}\left(D_{r^i}\|D^*\right)+1}, \quad (8)$$

where $d_{KL}(D_1\|D_2) = \sum_{j} D_1(j) \log_e \frac{D_1(j)}{D_2(j)}$ is the KL-divergence of distribution $D_1$ with respect to distribution $D_2$ and $Z = \sum_{i} \frac{1}{\log_2(i+1)}$. We add 1 to $d_{KL}$ which does not affect the trend of KL-divergence to counteract the divide-by-zero problem. Meanwhile, this has the nice property that the nDRKL is within range (0,1] with optimal value 1 at rank 1 and KL-divergence being zero. Contrary to nDKL which is the lower score the better fairness, the higher score of nDRKL means the better fairness. Meanwhile, nDRKL discourages bias at top rank positions by encouraging lower KL-divergence at higher ranks. nDRKL reports the average performance by averaging the scores over all queries.

## 5.1.4 | Results

The experiment results were reported in Tables 1 and 2. All algorithms on different datasets are statistically significant with respect to *p*-value .001. As mentioned before, the gold standard relevance judgments were used in the DetCons algorithm on ClueWeb09 data. We observed that there were enough relevant documents for each query such that the documents chosen in the top-$k$ ranking were all relevant. Consequently, the nDCG and RBP scores were both 1 for the DetCons on ClueWeb09 experiment.

Across different datasets, MovieLens had the worst utility (Table 1; nDCG and RBP) scores due to its extreme sparsity. FAIR was able to reflect this with the lowest score. Considering all the four utility and fairness scores, ClueWeb09 had slightly better relevance scores than Google Search, yet much worse fairness (Table 2; KL) scores. This was captured with higher FAIR scores on Google and Google+DetCons than on ClueWeb09. For each dataset and the given ranking algorithm, as $k$ increased, all utility and fairness metric scores improved. FAIR successfully reflected this increase. On the same datasets with different ranking algorithms, that is, Google versus Google DetCons, FAIR highlighted the dimension that had a major difference (results on fairness metrics from Table 2 are more different than results on utility metrics from Table 1). From Tables 1 and 2, we see that FAIR was able to highlight cases where both utility and fairness were good, as well as impose higher weights on the dimension that had more distinctive performances (e.g., ClueWeb09 vs. MovieLens on utility metrics). To further understand how FAIR correlates with each of the utility and fairness metrics, we conducted the correlation analysis between FAIR against other metrics. The results were summarized in Table 3. We did not report on nDCG or RBP on ClueWeb09 since both scores were 1 on this dataset. From Table 3, we see that FAIR had higher correlation with nDCG and RBP as the rank k increased. This indicated that at the top ranking positions, FAIR was more different from existing metrics where both utility and fairness were important. In other words, when we try

**TABLE 1** Summary of the top-$k$ results with utility-based metrics on three different datasets

| Datasets | nDCG | | | RBP | | |
| --- | --- | --- | --- | --- | --- | --- |
| | $k = 10$ | $k = 20$ | $k = 50$ | $k = 10$ | $k = 20$ | $k = 50$ |
| Google | s | 0.8931 | 0.9141 | 0.9754 | 0.9763 | 0.9764 |
| Google+DetCons | 0.8984 | 0.9200 | 0.9303 | 0.9692 | 0.9702 | 0.9702 |
| ClueWeb09 | 1.0000 | 1.0000 | 1.0000 | 1.0000 | 1.0000 | 1.0000 |
| MovieLens | 0.0229 | 0.0331 | 0.0647 | 0.0157 | 0.0157 | 0.0157 |

TABLE 2 Summary of the top-$k$ results with fairness-based metrics on three different datasets

|  | FAIR | | | KL | | | nDRKL | | |
| --- | --- | --- | --- | --- | --- | --- | --- | --- | --- |
| Datasets | $k = 10$ | $k = 20$ | $k = 50$ | $k = 10$ | $k = 20$ | $k = 50$ | $k = 10$ | $k = 20$ | $k = 50$ |
| Google | 0.6653 | 0.6878 | 0.7077 | 0.0915 | 0.0563 | 0.0091 | 0.8148 | 0.8593 | 0.9110 |
| Google+DetCons | 0.7705 | 0.7938 | 0.8041 | 0.0053 | 0.0009 | 0.0002 | 0.9092 | 0.9405 | 0.9673 |
| ClueWeb09 | 0.6001 | 0.6113 | 0.6436 | 0.1646 | 0.1438 | 0.1284 | 0.8333 | 0.8546 | 0.8761 |
| MovieLens | 0.0178 | 0.0275 | 0.0581 | 0.0730 | 0.0459 | 0.0308 | 0.8204 | 0.8665 | 0.9121 |

TABLE 3 Summary of the correlation between FAIR and other four metrics, that is, nDCG, RBP, KL, and nDRKL

|  | FAIR-nDCG | | FAIR-RBP | | FAIR-KL | | FAIR-nDRKL | |
| --- | --- | --- | --- | --- | --- | --- | --- | --- |
| Datasets | Pearson | Spearman | Pearson | Spearman | Pearson | Spearman | Pearson | Spearman |
| $K = 10$ | | | | | | | | |
| Google | 0.1907 | 0.4909 | 0.3396 | 0.8060** | −0.4351 | −0.8060** | 0.6234 | 0.8060** |
| Google+DetCons | 0.0057 | 0.5515 | 0.6649 | 0.9515*** | −0.5641 | −0.9515*** | 0.8137** | 0.9515*** |
| ClueWeb09 | – | – | – | – | −0.6552* | −0.1393 | 0.9968*** | 0.9999*** |
| MovieLens | 0.9809*** | 0.9878*** | 0.7881** | 0.9999*** | −0.8412** | −0.9999*** | 0.9527*** | 0.9999*** |
| $K = 20$ | | | | | | | | |
| Google | 0.3254 | 0.7939*** | 0.5110 | 0.9759*** | −0.6848*** | −0.9759*** | 0.8700*** | 0.9759*** |
| Google+DetCons | 0.3488 | 0.7563*** | 0.6864*** | 0.9939*** | −0.6337** | −0.9909*** | 0.9113*** | 0.9939*** |
| ClueWeb09 | – | – | – | – | −0.7157*** | −0.5067* | 0.9968*** | 1.0000*** |
| MovieLens | 0.9963*** | 0.9984*** | 0.5523* | 1.0000*** | −0.6675** | −1.000*** | 0.8734*** | 1.0000*** |
| $K = 50$ | | | | | | | | |
| Google | 0.6832*** | 0.9018*** | 0.5267*** | 0.9984*** | −0.7915*** | −0.9981*** | 0.9534*** | 0.9984*** |
| Google+DetCons | 0.6567*** | 0.8908*** | 0.6697*** | 0.9996*** | −0.6443*** | −0.9857*** | 0.9547*** | 0.9996*** |
| ClueWeb09 | – | – | – | – | −0.8115*** | −0.9005*** | 0.9931*** | 0.9999*** |
| MovieLens | 0.9997*** | 0.9999*** | 0.3548* | 0.9999*** | −0.5205*** | −0.9973*** | 0.8113*** | 0.9999*** |

*Significance level ($p < .05$); **Significance level ($p < .01$); ***Significance level ($p < .001$).

to capture both utility and fairness, FAIR metric differs the most from individual utility metrics as well as individual fairness metrics at top ranking positions. As k increased, the reward of selecting relevant and fair items diminished due to the large position discount. This indicated that FAIR was monotonically yet nonlinearly correlated with all metrics.

## 5.2 | Case study on FAIR $\varepsilon$-greedy algorithm

We have shown how FAIR correlates with other metrics. In this subsection, we present a case study on the Google Search data to: (1) further investigate how FAIR changes with ranking algorithms; and (2) how FAIR $\varepsilon$-greedy algorithm performs compared to the algorithms that optimize other metrics.

### 5.2.1 | Algorithms and metrics

In this case study, we evaluated the following algorithms: Google (Google's original ranking), DetCons, the greedy algorithm that approximates the ideal $\alpha$-nDCG (IDCG), and our proposed FAIR $\varepsilon$-greedy. When $\varepsilon = 1$, the FAIR $\varepsilon$-greedy served as the algorithm that minimized KL-divergence at each rank position. The IDCG and the corresponding rankings were computed by a greedy approach that maximized the gain $G[i]$ (Equation 4) at each rank. We examined a different set of metrics for the purpose of this case study: $\alpha$-nDCG, KL, nDRKL, minSkew, maxSkew, and feasibility (Geyik et al., 2019). $\alpha$-nDCG is an example of utility metrics that measure the system's capability of retrieving relevant and diverse results. MinSkew and maxSkew correspond to the best case and the worst case fairness performance of an algorithm by taking the minimum and maximum skewness among all queries.

**TABLE 4** MaxSkew results on Google Search data

|  |  | Google | DetCons | IDCG | 0-greedy | 0.3-greedy | 0.5-greedy | 0.7-greedy | 1-greedy |
|---|---|---|---|---|---|---|---|---|---|
| NoRJ | $k = 10$ | 3.4702 | 2.3964 | 4.3340 | 4.2341 | 4.1002 | 3.9241 | 3.5521 | 2.7246 |
|  | $k = 20$ | 3.8622 | 3.0445 | 6.2146 | 4.4067 | 4.4046 | 4.3357 | 4.0079 | 3.9120 |
|  | $k = 50$ | 4.2732 | 3.9318 | 5.9435 | 4.4427 | 4.4427 | 4.4422 | 4.4064 | 4.0110 |
| RJ | $k = 10$ | 3.4568 | 2.3964 | 5.5215 | 4.0604 | 3.9290 | 3.8993 | 3.7992 | 2.5257 |
|  | $k = 20$ | 3.8622 | 3.0445 | 6.2146 | 4.4067 | 4.4035 | 4.3426 | 4.0066 | 3.9120 |
|  | $k = 50$ | 4.2732 | 3.9318 | 7.1309 | 4.4427 | 4.4427 | 4.4425 | 4.4066 | 3.9730 |

*Note*: $\varepsilon$-greedy stands for FAIR $\varepsilon$-greedy algorithm.

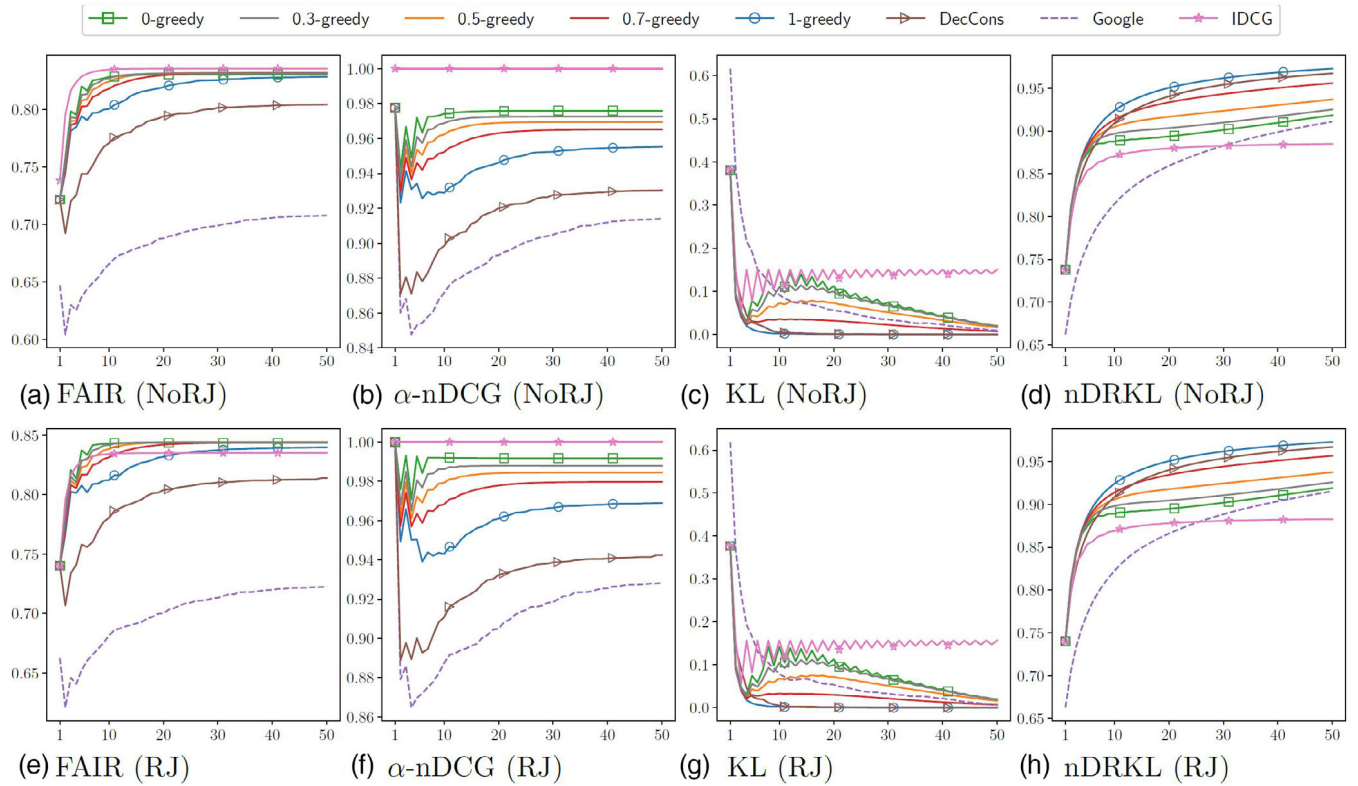

**FIGURE 1** Results on Google Search data. *x*-axis is rank position; *y*-axis is metric score

Feasibility reveals to what extend fairness is guaranteed, that is, at how many rank positions and for how many queries among all queries, the fairness constraints are violated.

### 5.2.2 | Assumption on availability of relevance judgment

For comparison, we considered two scenarios for the Google Search data. The first scenario (referred as NoRJ) assumed that we did not have access to the relevance judgment, and were only provided with Google's original ranking for each query. The second scenario was that we were provided with the gold standard relevance judgment (referred as RJ). In both cases, re-ranked results were evaluated with the gold standard relevance judgment.

### 5.2.3 | Results

The minSkew scores were all 0 s. The maxSkew scores were reported in Table 4. The Google and IDCG were not feasible, and DetCons and FAIR $\varepsilon$-greedy were only feasible up to rank 20. The FAIR, $\alpha$-nDCG, KL and nDRKL scores were plotted in Figure 1. We eliminated the term FAIR from FAIR $\varepsilon$-greedy algorithms in the legends of the figure. The results for FAIR $\varepsilon$-greedy were obtained by averaging over 1,000 runs to account for the randomization of this algorithm.

The observation on skewness and feasibility demonstrated that, for some strategies that simply optimized for utility, there could be serious algorithmic bias as in the case of IDCG. On the other hand, the results of DetCons and our FAIR $\varepsilon$-greedy algorithm suggested that, when carefully designed, it was possible for an algorithm to avoid or at least reduce the bias while optimizing for utility. From Figure 1, we see that for all metrics that accounted for the position bias (FAIR, $\alpha$-nDCG and nDRKL), our algorithms outperformed the Google baseline, even when relevance judgment was not available (NoRJ). The comparisons between FAIR $\varepsilon$-greedy and other algorithms showed that we could increase utility and reduce bias at the same time.

Through the case study, we can further validate the contribution of our proposed FAIR metric. Figure 1 showed that FAIR rewarded the algorithms that were good in both fairness and utility. Ideally, the high FAIR score should correspond to high user satisfaction. Diversity- and novelty-based relevance metrics have been found to correlate well with user satisfaction in many IR applications, especially for web applications (e.g., Huffman & Hochster, 2007). By design (Equation 6), FAIR strikes a balance between diversity- and novelty-based relevance and fairness while trying to optimize the combined scores of both. In Figure 1a,e, we can indeed observe some positive correlation with $\alpha$-nDCG and nDRKL. Existing diversity and novelty metrics such as $\alpha$-nDCG do not account for fairness, thus a higher utility score does not necessarily mean a higher FAIR score (e.g., IDCG). Similarly, a higher score of fairness does not necessarily imply a higher FAIR score (e.g., DetCons). However, if the utility score is high and the bias is low, we must have a high FAIR score (e.g., $\varepsilon$-greedy vs. Google ranking). A low FAIR score indicates either the utility is low or the bias is high, or both, as can be seen from the performance of Google and DetCons. This again highlighted the capability of FAIR to balance between utility and fairness as an integrated evaluation metric.

## 6 | DISCUSSION

We must point out that the new fairness-aware metric requires further testing through various datasets and applications in order to show whether it truly captures the user models in the retrieval and recommendation tasks. There are limitations in the current design of the metric that may be improved in the future work.

For example, FAIR is only valid for effectiveness metrics defined as a sum over the ranking, so they can be plugged in Equation (1). This can be addressed by removing the outside sum, taking the utility metric as the entire numerator with the fairness metric as the entire denominator. Additionally, the normalizer in the current FAIR does not account for the case where perfect fairness cannot be achieved. Therefore, a ranking that is different from the desired distribution always gets punished in the FAIR score. This should be addressed by first implementing an algorithm that minimizes the difference from the desired distribution (such as our FAIR 0-greedy algorithm), and then including the fairness score of that algorithm in the normalizer. Moreover, FAIR is not proposed to be a "one size fits all" metric. When there are interdependences between utility and fairness metrics, such interdependences must be carefully eliminated or addressed in other forms to make the integration meaningful. When there exist known trade-offs between utility and fairness, it is better to directly model the trade-offs as a metric, rather than treating them as factors assuming no knowledge about their relationship. We must note we cannot and should not compare two FAIR scores that are computed using different definitions of utility and fairness, just as we should not compare an nDCG score with an RBP score. Lastly, there are cases where two algorithms have the same FAIR score. One algorithm has a significantly higher utility score yet an extremely low fairness score, the other one has the opposite performance in terms of utility and fairness. In this case, the FAIR metric cannot tell these two algorithms apart. This could be improved by tuning the weights between utility and fairness. However, this is an inherent drawback of evaluating with a single integrated metric. We also acknowledge that, with our current design of the metric, there are limited metrics that can be plugged in, especially those variants of precision, diversity- and novelty-based metrics. Handling with unbounded relevance or fairness metrics requires future work.

It is worth mentioning that we use KL-divergence in our example FAIR, and in evaluation metrics, but this should not be confused as evaluating a metric with a metric. What we are doing here is designing an algorithm that optimizes for some combination of two metrics/goals. Then we evaluate this algorithm on the individual metric to see how we capture this metric using this algorithm. This is similar to encoding an optimization dimension as part of the optimization function and evaluating how the new algorithm based on the function performs w.r.t. the dimension. It may or may not be the case that the algorithm does not capture this dimension well when mixing with other optimization dimensions.

While we believe that fairness and lack of bias are different (more like equity vs. equality), in this article, we use the terms interchangeably to simplify the discussion around fairness and bias. We refer interested readers to Gao and Shah (2021) for differences and connections between fairness and bias. FAIR is designed with a focus on group fairness rather than individual fairness. The question with fairness often is "fair to who?" As we are primarily focused on end-user-based metrics, we need to consider the other side of this story—that is, information sources or topics. The notion of group fairness allows us to consider diversity and representation of those topics and perspectives as well.

In this article, we are working under the "classic assumption" that relevance is one way to represent *utility*. However, given the rising attention on biases prevalent in IR systems, what is considered as utility becomes an important topic that is worth debating. Utility can cover many aspects of a system, even fairness itself could be a measure of utility. While replacing utility with "relevance" will make it an easier task, we believe keeping the discussion of utility in this thread allows for future generalizations and debate over what really is utility. The results remain unaffected, but the discourse can change.

## 7 | CONCLUSIONS

In this article, we proposed a novel fairness-aware metric that was specifically designed to evaluate the performance of fairness ranking algorithms. This new metric combined the essential factors in IR—relevance, diversity, and novelty—discounted by the position bias and diversity bias. Through experiments on various datasets and a detailed case study, we highlighted the contribution of our FAIR metric. FAIR was not only able to reflect both the utility and fairness, but also the balance between the two, which none of the existing utility or fairness metrics could achieve. Based on this new metric, we developed an effective algorithm to jointly optimize both utility and fairness. We showed that, compared to the baseline algorithm, the FAIR-based optimization algorithm was able to achieve better utility as well as fairness. This further demonstrated the usefulness of an integrated metric as FAIR.

We believe our work opens up a new direction of pursuing a computationally feasible metric for evaluating and implementing fairness-aware IR systems. Ultimately, we aim to show how fairness and utility, even with our imperfect notions and measures, could be studied under the same lens of optimizing system performance and user satisfaction. We believe this could help system designers and perhaps even policy makers. Rather than prescribing a specific combination or trade-offs of fairness and utility, we are focusing on creating tools and metrics for designers and policy makers to make decisions. Integrating both of these measures into one allows for a better understanding of how overall system performance is affected by individual factors and how one may want to optimize for integrated metrics rather than individual components.

## ORCID

*Ruoyuan Gao*  https://orcid.org/0000-0002-8784-4171

## ENDNOTES

[1] When the context is clear, we will write $d_{KL}\left(D_{r^i} \middle\| D^*\right)$ as $d_{KL}$ for short.

[2] https://trec.nist.gov/data/web09.html